 \documentclass[final,3p,times,authoryear]{elsarticle}
\usepackage{amssymb}
\usepackage{amsmath}
\usepackage{caption}
\usepackage{subfigure}
\usepackage[switch,mathlines]{lineno}
\journal{New Astronomy}

\begin{document}

\begin{frontmatter}

%% Title, authors and addresses

%% use the tnoteref command within \title for footnotes;
%% use the tnotetext command for theassociated footnote;
%% use the fnref command within \author or \affiliation for footnotes;
%% use the fntext command for theassociated footnote;
%% use the corref command within \author for corresponding author footnotes;
%% use the cortext command for theassociated footnote;
%% use the ead command for the email address,
%% and the form \ead[url] for the home page:
%% \title{Title\tnoteref{label1}}
%% \tnotetext[label1]{}
%% \author{Name\corref{cor1}\fnref{label2}}
%% \ead{email address}
%% \ead[url]{home page}
%% \fntext[label2]{}
%% \cortext[cor1]{}
%% \affiliation{organization={},
%%            addressline={}, 
%%            city={},
%%            postcode={}, 
%%            state={},
%%            country={}}
%% \fntext[label3]{}

\title{Photometric analysis of the dwarf nova SU UMa based on TESS}

%% use optional labels to link authors explicitly to addresses:
%% \author[label1,label2]{}
%% \affiliation[label1]{organization={},
%%             addressline={},
%%             city={},
%%             postcode={},
%%             state={},
%%             country={}}
%%
%% \affiliation[label2]{organization={},
%%             addressline={},
%%             city={},
%%             postcode={},
%%             state={},
%%             country={}}

\author[1,2,3]{ Wei Liu\corref{mycorrespondingauthor}}
\cortext[mycorrespondingauthor]{Corresponding author}
\ead{liuwei@ynao.ac.cn}
\author[1,2,3]{Xiang-Dong Shi}
\author[4]{Xiao-Hui Fang}
\author[4]{Qi-Shan Wang}
\address[1]{Yunnan Observatories, Chinese Academy of Sciences, P.O. Box 110, Kunming 650216, P.R. China}
\address[2]{University of Chinese Academy of Sciences, No.1 Yanqihu East Rd, Huairou District, Beijing 101408, P.R. China}
\address[3]{Key Laboratory of the Structure and Evolution of Celestial Objects, Chinese Academy of Sciences, P.O. Box 110, 650216 Kunming, P.R. China}
\address[4]{School of Mathematics, Physics and Finance, Anhui Polytechnic University, Wuhu 241000, P.R. China}

\begin{abstract}

We report the photometric analysis of SU UMa based on the observations of 
the Transiting Exoplanet Survey Satellite (TESS). TESS has released a large amount of data, which contains the light curves of a complete superoutburst and three normal outbursts of SU UMa. Based on the observations, the evolution of superhumps during the superoutburst was analyzed. By using the O - C method, the three stages were determined, and the superhump period for each stage was calculated. The mass ratio $q=0.137 (1)$ was estimated based on the stage A superhump method.  A periodic oscillation with a period of 2.17(9) days was found in the superhump minimum O - C diagram, which is related to the precession of the accretion disk. We investigated the frequency information of quasiperiodic oscillations (QPOs) in the light curves. An ultralong period QPO of 0.18 days at the time of normal outburst and two ultralong periods QPOs of 0.19 days and 0.285 days in quiescence were found. No QPO was found in the superoutburst.

\end{abstract}

%%Graphical abstract
%\begin{graphicalabstract}
%\includegraphics{grabs}
%\end{graphicalabstract}

%%Research highlights
%\begin{highlights}
%\item Research highlight 1
%\item Research highlight 2
%\end{highlights}

\begin{keyword}
%% keywords here, in the form: keyword \sep keyword, up to a maximum of 6 keywords
photometric \sep dwarf novae \sep binary star \sep stars: individual (SU UMa) 

%% PACS codes here, in the form: \PACS code \sep code

%% MSC codes here, in the form: \MSC code \sep code
%% or \MSC[2008] code \sep code (2000 is the default)

\end{keyword}

\end{frontmatter}

%\tableofcontents
%\linenumbers

%% main text
%%\onecolumn
\section{Introduction}
\label{introduction}

Cataclysmic variables (CVs) are semicontact binary systems that contain a white dwarf and a red dwarf \citep{2003cvs..book.....W}. Dwarf novae (DNe) is a subtype of nonmagnetic CVs that is currently the most abundant subgroup. The red dwarf fills its Roche lobe and transfers mass to the white dwarf via the inner Lagrange point. The transferred mass will form a disk around the white dwarf. In magnetic CVs, due to the strong magnetic field of the white dwarf, the transferred mass is trapped and moves along the magnetic field lines to reach the poles of the white dwarf, so a complete accretion disk will not form.

SU Ursae Majoris (SU UMa) is the prototype star for a subgroup of DNe with an orbital period of $0.07635(4)$ days  \citep{1986ApJ...309..721T}.  It was thought to be a noneclipsing binary with  V magnitudes ranging from 11.3 to 15.7\citep{2003A&A...404..301R}. The mass ratio and the mass of its two components are not yet known. Based on the ultraviolet spectra of the International Ultraviolet Explorer, \citet{2017NewA...52..122Z} estimated an accretion rate of approximately $9.8\times10^{-13} M_{\odot}yr^{-1}$.  

SU UMa-type dwarf novae (SU UMa stars) with orbital periods below the period gap (2-3 h) show superoutbursts and normal outbursts. The outburst cycles of SU UMa stars are indefinite. Superoutbursts last approximately ten days and are longer than normal outbursts but have less frequency. The mechanism of superoutbursts is considered to be the combination of disk instability and tidal instability, which is different from normal outbursts\citep{1989PASJ...41.1005O}. A superoutburst is usually triggered by a normal outburst. The outburst mechanism of normal outbursts in SU UMa stars is the same as that of U Gem stars, which are described by disk instability model  \citep{1974PASJ...26..429O,2001NewAR..45..449L} or mass
transfer instability model \citep{1975MNRAS.171..311B}. The tidal instability of accretion disks is caused by a resonance between the orbiting secondary star and disk particle orbits with a 3:1 period ratio \citep{1990PASJ...42..135H}. 

Superhumps are periodic brightness variations in SU UMa stars with periods a few percent longer than the orbital period. They generally appear during the plateau of superoutbursts, which originate from the resonance between the elliptical disk and binary orbit \citep{1988MNRAS.232...35W}. According to the envolution of the superhump period ($P_{sh}$) of SU UMa stars, \citet{2009PASJ...61S.395K} divided superhumps into three distinct stages: a longer superhump period in the early envolutionary stage (stage A), varying periods in the middle stage (stage B), and a shorter superhump in the final stage (stage C). They later developed the stage A superhump method to estimate the mass ratio of binary stars based on their superhump period of stage A and orbital period \citep{2013PASJ...65..115K}. Based on the photometric observations of the April 1989 superoutburst of SU UMa, \citet{1990AJ....100..226U} discovered superhumps with a period of 113.4 min (0.07875 d) during the late stages of the superoutburst. The January 2010  superoutburst was observed by \citet{2010PASJ...62.1525K}, but only a mean period of stage B of 0.07907(3) d was obtained. Unlike superhumps, negative superhumps are periodic light modulations with a period a few percent shorter than the orbital period, which has not been detected in SU UMa. They originate from a retrogradely precessing tilted accretion disk and are produced by the transit of the accretion steam impact spot across the face of the tilted disk \citep{2007ApJ...661.1042W}. Superhumps are slightly similar to negative superhumps, but the mechanisms of their generation are completely different.
\citet{2012PASJ...64L...5I} discovered superhumps in a normal outburst in January 2012.

The intervals of superoutbursts (supercycles) are longer than those of normal outbursts. SU UMa exhibits normal outbursts at intervals generally ranging from 5–33 days, and the supercycle is approximately 160 days \citep{1979AJ.....84..804P,1984PASP...96..988S,2013PASJ...65...87I}.
Superoutbursts are unpredictable and last from ten to dozens of days, so it is difficult to obtain their complete light curve. Supercycles can also reflect the material transfer rate from the companion to the accretion disk. 

Quasiperiodic oscillations (QPOs) are a common phenomenon in the light curves of CVs \citep{1977ApJ...214..144P}. They are short-time-scale modulations in light curves that refer to the luminosity variation with time ranging from 50 to 1000 seconds  \citep{2002MNRAS.333..411W}. 
QPOs are difficult to recognize because of their low coherence and short coherence time. 
In previous studies, based on ground-based telescopes, the QPOs during quiescence and normal outbursts in SU UMa were discussed.  A QPO with period of  $\sim 0.0481(4)$ days during the rising stage of a normal outburst was found \citep{2013PASJ...65...87I}.
Since no superoutburst was observed, their discussion was not comprehensive enough. 

In this paper, we present light curve studies of SU UMa. Section 2 introduces the observations and data of SU UMa. Section 3 shows the analysis of the evolution of the superhump during the superoutburst. Based on this, we constrained some system parameters of this system. The O - C analysis of the superhump minimum is presented in Section 4. QPOs and their variations are investigated and discussed in Section 5. Section 6 contains a conclusion.

\section{The observations and data}

\begin{figure*}
    \centering
	\includegraphics[width=\linewidth]{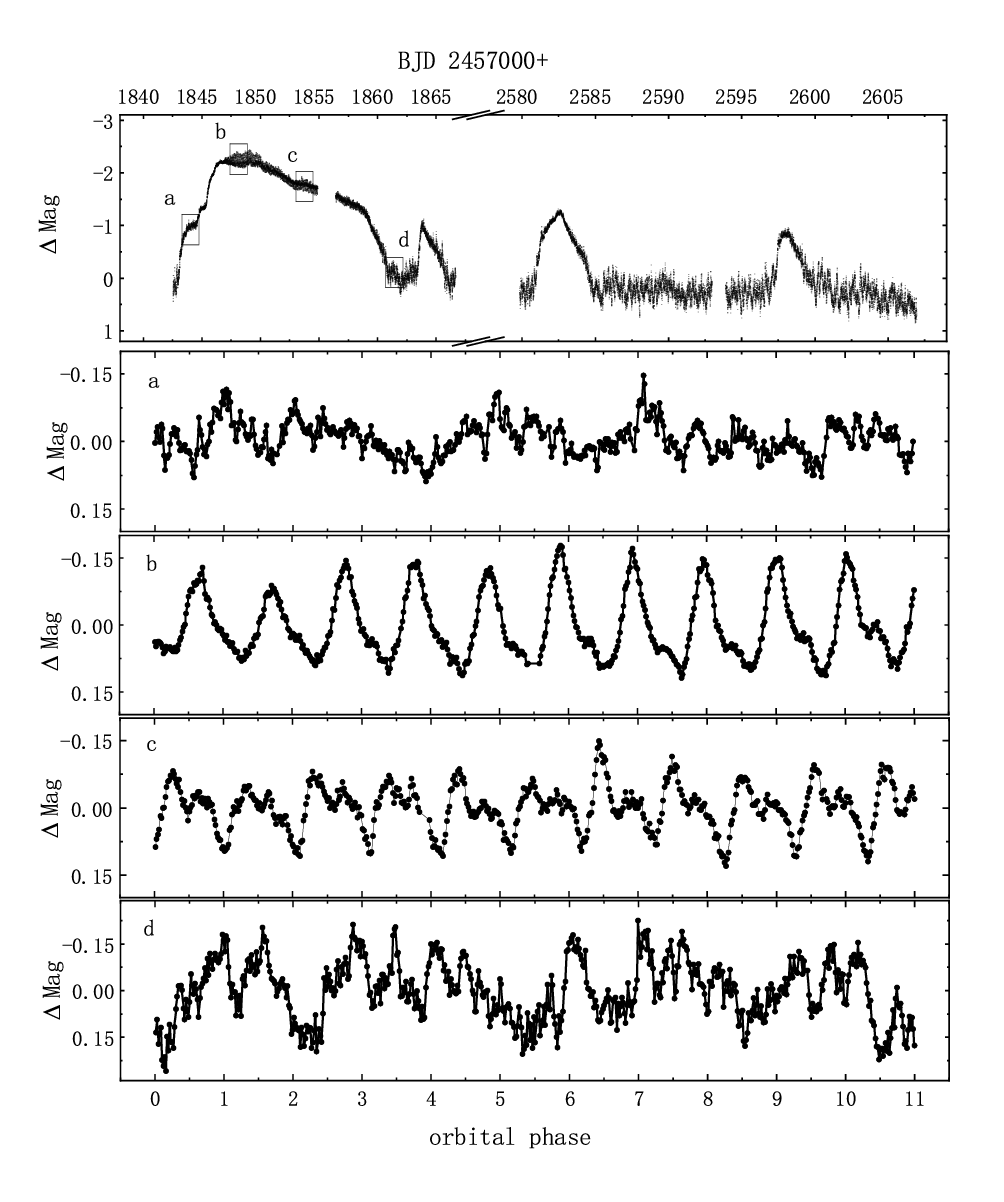}
    \caption{The top panel shows light curves from TESS, which are shown in relative magnitude and Barycentric Julian Date  (BJD). The relative magnitude was calculated by the formula $\Delta mag = -2.5log_{10}(sap\_flux) + 6$. Panels a, b, c, and d show the enlargement of the data in each rectangle marked in the top panel after the removal of global trend . The orbital phases were calculated by the orbital period 0.07635 .}
    \label{fig:overall}
\end{figure*}

The data were taken from the Transiting Exoplanet Survey Satellite (TESS) \citep{ricker2015transiting}.  The main mission of TESS is to search for exoplanets via transit detection. Many outbursts of DNe were observed. Continuous monitoring allows us to study the properties of superhumps better than ground-based telescopes. The photometric observations by TESS were described in barycentric julian date (BJD) and the flux of electron/s. The data were downloaded from the Mikulski Archive for Space Telescopes (MAST) data archive
\footnote{https://mast.stsci.edu/portal/Mashup/Clients/Mast/Portal.html}.
Two sectors' data were obtained from 25 December 2019 to 19 January 2020 and from 31 December 2021 to 27 January 2022 with a cadence time of 120 seconds. Every sector has a one-day gap for transferring data back. The detectors attached to TESS are sensitive from 600 nm to 1000 nm.   SAP data were adopted in this work. To identify the amplitude of outbursts and superhumps more easily, we transferred the SAP flux to the relative magnitude and set the magnitude in quiescence to 0. All light curves of SU UMa are shown in the top panel of Fig.  \ref{fig:overall}. The light curves of a complete superoutburst and 3 normal outbursts are observed, and two shoulders occur at the presuperoutburst. The interval between the peaks of the later two consecutive normal outbursts is 15 days.

\section{Analysis of Superhump maxima}
\begin{figure*}

	\includegraphics{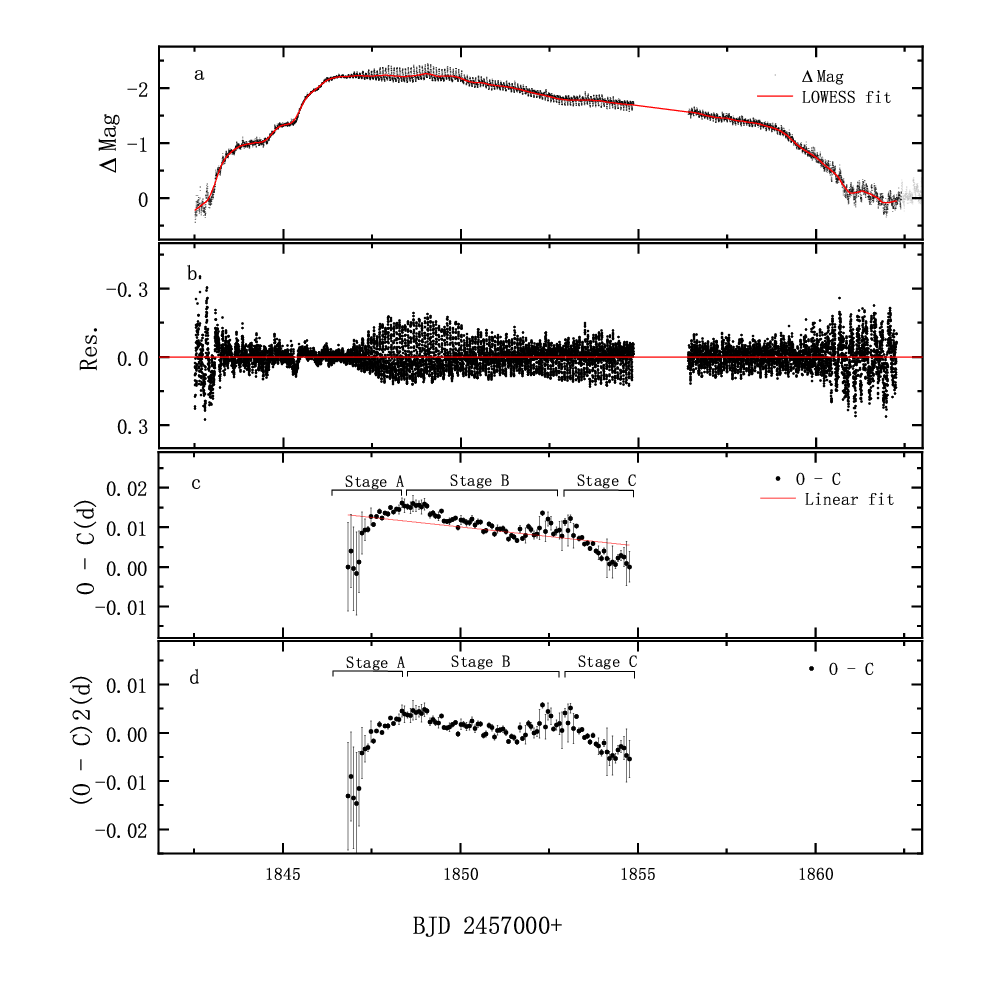}
    \caption{The panel a shows light curve of the superoutburst, and the red line represents the superoutburst trend. The relative magnitude was caculated by the formula $\Delta mag = -2.5log_{10}(sap\_flux) + 6$.  Panel b shows the  detrended light curve, we can see the amplitudes variation from the panel directly. Panel c shows the O - C diagram of the superhumps and the linear fit of the O - C data. Panel d shows the revised O - C diagram of the superhumps. }
    \label{fig:SO}

\end{figure*}

 Superhumps can always be seen during superoutbursts of SU UMa stars.  The 2020 January superoutburst was completely observed by TESS except for a one-day gap (see Fig. \ref{fig:SO}). The superoutburst lasts approximately 19.5 days with an amplitude of approximately 2.5 magnitude, which is 3 times longer than a normal outburst. We removed the superoutburst trend in the light curve by using the locally weighted scatter plot smoothing (LOWESS) method.  
According to the detrended light curves, the changes in amplitude of superhumps can be viewed directly (as panel b of Fig. \ref{fig:SO}). The superhump takes approximately one day from its appearance to full amplitude , remains for approximately two days, and then begins to decrease and gradually disappear.

\begin{table*}
	\centering
	\caption{Superhump period of SU UMa}
	\label{tab:example_table}
	\begin{tabular}{lcccc} % four columns, alignment for each
		\hline
		Stage & E & BJD-2457000 &  Period (d) & error(d)\\
		\hline
		A & 0-20 & 1846.80-1848.45 & 0.08013   & 0.00004\\
		B & 20-72 & 1848.45-1852.55 & 0.07920  & 0.00002\\
		C & 73-100 & 1852.55-1854.86 & 0.07886 & 0.00004\\
		\hline
	\end{tabular}
\end{table*}

Using the quadratic polynomial fitting method, we obtained the light maximum times of superhumps. only the light maximum times from BJD 2458846 to 2458854 were taken because of the low signal-to-noise ratio at the beginning and end of the superoutburst. Based on the generalized Lomb-Scargle method \citep{1976Ap&SS..39..447L,1982ApJ...263..835S,1989ApJ...338..277P}, the mean period of the superhumps was derived as 0.079281(1) days. The ephemeris of superhumps can be written as 

\begin{equation}
    T_{max} = BJD2458846.8326(11)+0.079281(1)\times E,
    \label{Eq:1}
\end{equation}
where BJD2458846.8326(11) is the initial epoch, and E is the cycle number of superhumps. According to the ephemeris and the light maximum times, the O - C values were obtained. 

\begin{figure*}
	% To include a figure from a file named example.*
	% Allowable file formats are eps or ps if compiling using latex
	% or pdf, png, jpg if compiling using pdflatex
    \includegraphics[width=\linewidth]{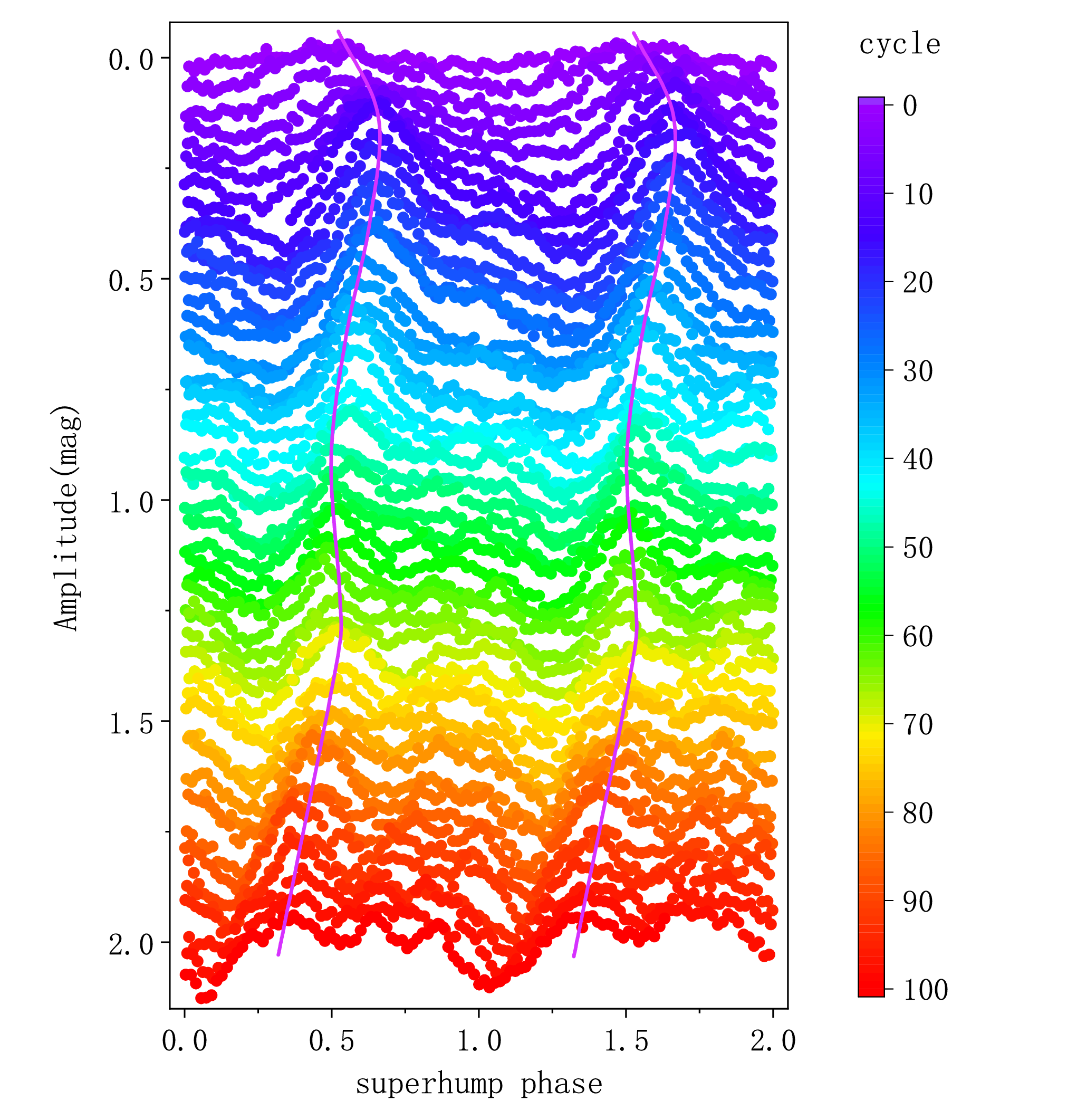}
    \caption{The superhump phase is calculated by equation \ref{Eq:1} and shifted afterward by 0.5 phases. The color bar represents the superhump cycle. Two purple lines show the trend of the maximum with cycles. The trend corresponds to stages A, B, and C.}
    \label{fig:superhumpzd}
\end{figure*}

The O - C diagram is shown in panel c of Fig. \ref{fig:SO}. After a linear correction, typical stages A, B, and C appear in panel d of Fig. \ref{fig:SO}. According to the linear fit, the mean superhump period was revised to 0.079164(9) days. Stages A, B, and C can be clearly distinguished from this O - C diagram (as identified in Fig.\ref{fig:SO}). Table \ref{tab:example_table} lists the information of each stage.  The evolution of the superhumps can also be seen in  Fig.\ref{fig:superhumpzd}. After the superhump reaches its full amplitude, it gradually shows a second hump after cycle 40, as seen in Fig.\ref{fig:superhumpzd} as well as in panel c of Fig.\ref{fig:overall}.
Two purple lines in Fig.\ref{fig:superhumpzd} show the trend of the light maximum phases with cycles. The trend corresponds to stages A, B, and C as the O - C distribution in panel d of Fig.\ref{fig:SO}. 

The stage A superhump method proposed by \citet{2013PASJ...65..115K}, a dynamical method that relies only on celestial mechanics, was adopted to calculate the mass ratio. 
Using the orbital period and the stage A superhump period, the fractional superhump excess in frequency is 

\begin{equation}
    \epsilon^{\ast} = 1 - \frac{P_{orb}}{P_{sh}},
\end{equation}
where $P_{orb}$ is orbital period and $P_{sh}$ is superhump period.
 For stage A, $P_{sh} = 0.08013(4)$, $\epsilon^{\ast} = 0.04717(4)$ . 
The dynamical precession rate, $\omega_{dyn}$ in the disk can be expressed by (see \citet{2016PASJ...68..107K}):

\begin{equation}
   \frac{\omega_{dyn}}{\omega_{orb}} =  \frac{q}{\sqrt{1+q}}[\frac{1}{4}\frac{1}{\sqrt{r}}b_{3/2}^{(1)} ]
\end{equation}
where $r$ is the dimensionless radius measured in units of binary separation $a$ and $q$ is the mass ratio (M2/M1). $\frac{1}{2}b_{s/2}^{(j)}$ is the Laplace coefficient 

\begin{equation}
   \frac{1}{2}b_{s/2}^{(j)} = \frac{1}{2\pi}\int_0^{2\pi}\frac{cos(j\phi)d\phi}{(1+r^2-2rcos\phi)^2}.
\end{equation}

During stage A, it is considered that the superhump wave is confined to the 3:1 resonance, and $r$ can be described as 

\begin{equation}
   r_{3:1} = 3^{(-2/3)}(1+q)^{-1/3}.
\end{equation}

This $\omega_{dyn}/\omega_{orb}$ is equal to the fractional superhump excess in frequency ($\epsilon^{\ast}$). According to these equations, the mass ratio of $q=0.137 (1)$ is derived.

\begin{figure*}
    \includegraphics{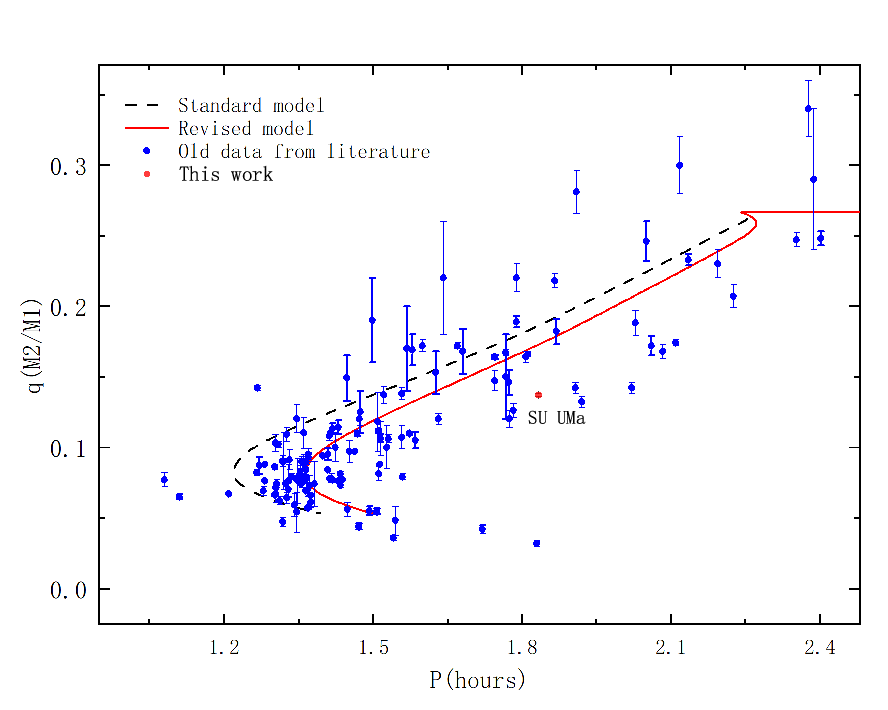}
    \caption{Mass ratio–period relation of CVs. The blue points shown are mass ratio and period form literature \citep{2022arXiv220102945K}. The data of the red dot is obtained in this work. The black dashed line shows the standard model for CV evolution. The red solid line shows the evolution of donor properties along the revised model track \citep{2011ApJS..194...28K}. }
    \label{fig:qpt}
\end{figure*}

The orbital period-mass ratio relation of CVs is shown in Fig.
\ref{fig:qpt}. The black dashed line and red line represent the standard and revised evolution tracks of CVs  \citep{2011ApJS..194...28K}.  The blue dots represent other CVs, the data of which are listed in \citet{2022arXiv220102945K}. As the mass ratio $q$ of SU UMa derived above, we put the data of SU UMa on this picture. As seen from the figure, the q-value is slightly lower than the model predictions. SU UMa is in the middle stage of the evolution of a short-period ($P_{orb}$ below the period gap) CV and evolving toward a shorter orbital period.  It will take a long time to evolve to its period minimum.

Based on the semiempirical donor tracks proposed by \citet{2011ApJS..194...28K} and the orbital period, the donor mass was estimated as $0.138(1)M_\odot$.  With $q = 0.137 (1)$ , the mass of the white dwarf is estimated to be $M_1 = 1 M_\odot$. 

\section{Analysis of Superhump minimum}
\begin{figure*}
	% To include a figure from a file named example.*
	% Allowable file formats are eps or ps if compiling using latex
	% or pdf, png, jpg if compiling using pdflatex
	\includegraphics{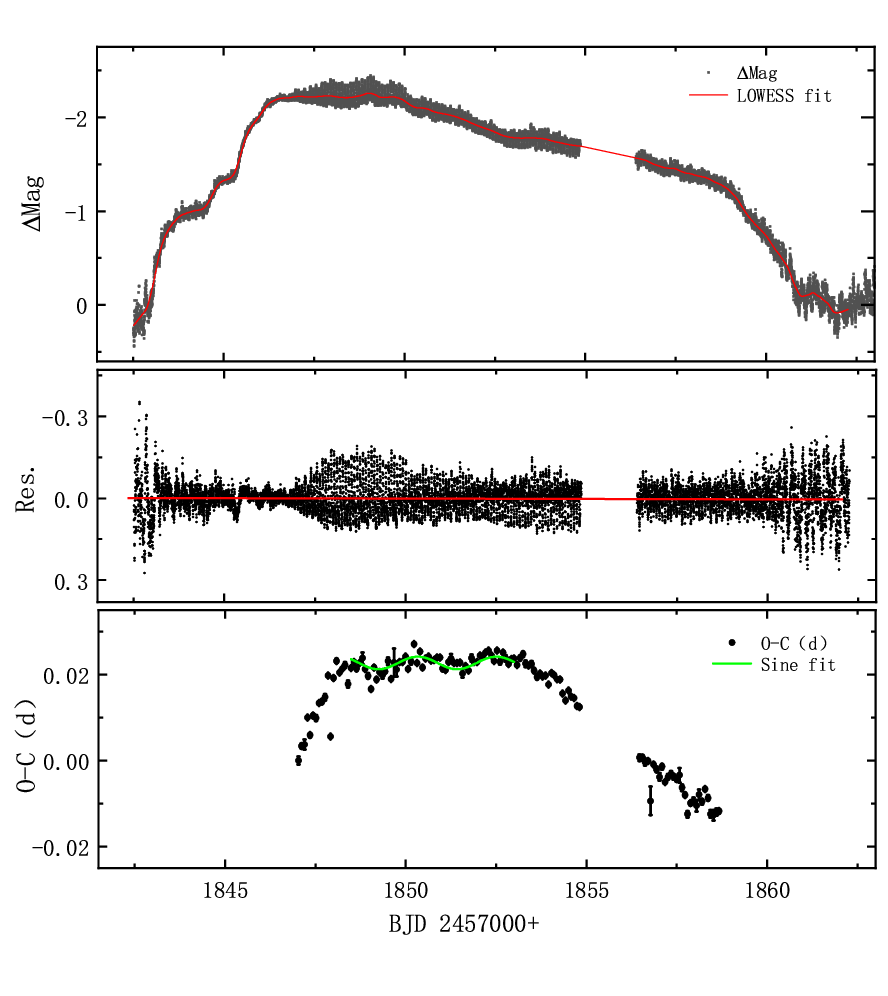}
    \caption{The top panel shows the compete superoutburst and the LOWESS fit of the light curve. The relative magnitude was caculated by the formula $\Delta mag = -2.5log_{10}(sap\_flux) + 6$. The middle panel shows the Residuals of the LOWESS fit, we can see the variation of the amplitude from the panel directly. The bottom panel shows the O - C analysis of the superhump minima, and the green line is the sine fit of the O - C variation.}
    \label{fig:eclipse}
\end{figure*}

 The superhumps show clear superhump minima in the plateau of the superoutburst from BJD 2458847 to 2458858.7, which last approximately 11.7 days and longer than the superhump maxima (8 days). They show a much higher signal-to-noise ratio (SNR) than the superhump maxima and disappear much later.  As shown in panel c of Fig. \ref{fig:overall}, the superhump minimum profile changes with the outburst, which rapidly disappears in the postsuperoutburst stage.  
A high SNR superhump minimum was selected as the initial epoch of the ephemeris. According to the quadratic polynomial fitting method, the times of superhump minima were obtained. We obtained a mean period of 0.079252 days by the light minima times. Taking the mean period, the ephemeris can be written as follows: 
\begin{equation}
    T_{min} = BJD2458847.03549(9)+0.079252\times E.
    \label{Eq:ephe2}
\end{equation}
From the equation, the O - C diagram was obtained, which is shown in the bottom panel of Fig. \ref{fig:eclipse}. 

The O - C diagram shows a downward parabolic shape, and a sinusoidal variation occurs from BJD 2458848.5 to 2458853. The superhump minima O - C diagram is similar to that of the superhump maxima; both are parabolic with an opening downward, except that the superhump minima O - C exhibits a distinct sinusoidal variation at the peak of the superoutburst. The variation lasts approximately $4.5$ days which is shorter than the duration of the superhumps. We used a sine function to fit the variation, and a period of 2.17(9) days was obtained. 

Precession of the accretion disk widely occurs during superoutbursts of SU UMa stars with a period of several days. The O - C variation occurs at the time of accretion disk precession, and the cycle is similar, which implies that the O - C variation period (2.17(9) days) may be due to the precession of the disk. To verify whether this variation is related to precession, we calculated the period of precession using the orbital period ($P_{orb} = 0.07635$) and the mean superhump period ($P_{sh} = 0.079164(9)$ days) given above. According to the formula,

\begin{equation}
    \frac{1}{P_{prec}} = \frac{1}{P_{orb}} - \frac{1}{P_{sh}},
    \label{Eq:prec}
\end{equation}
the period of precession is derived as $P_{prec} = 2.15 $ days, which is only 0.02 days shorter than the O - C variation. Considering the variation in the period of the superhump and precession , we believe that the 2.17(9)-day variation is related to the precession of the accretion disk. The O - C method may be a new tool to determine the period of precession.

\section{Investigation of quasiperiodic oscillations}
QPOs are obvious during outbursts in some CVs (e.g., EM Cyg \citep{2021MNRAS.505..677L} and HS 2325+8205 \citep{2023MNRAS.518.3901S}), but not obvious during quiescence. They are mostly on the order of tens of seconds to thousands of seconds. QPOs are presented in the light curves of SU UMa clearly, as shown in panel d in Fig. \ref{fig:overall}. In particular, it is obvious in quiescence but not in outbursts.

\begin{figure*}
    \includegraphics[width=\linewidth]{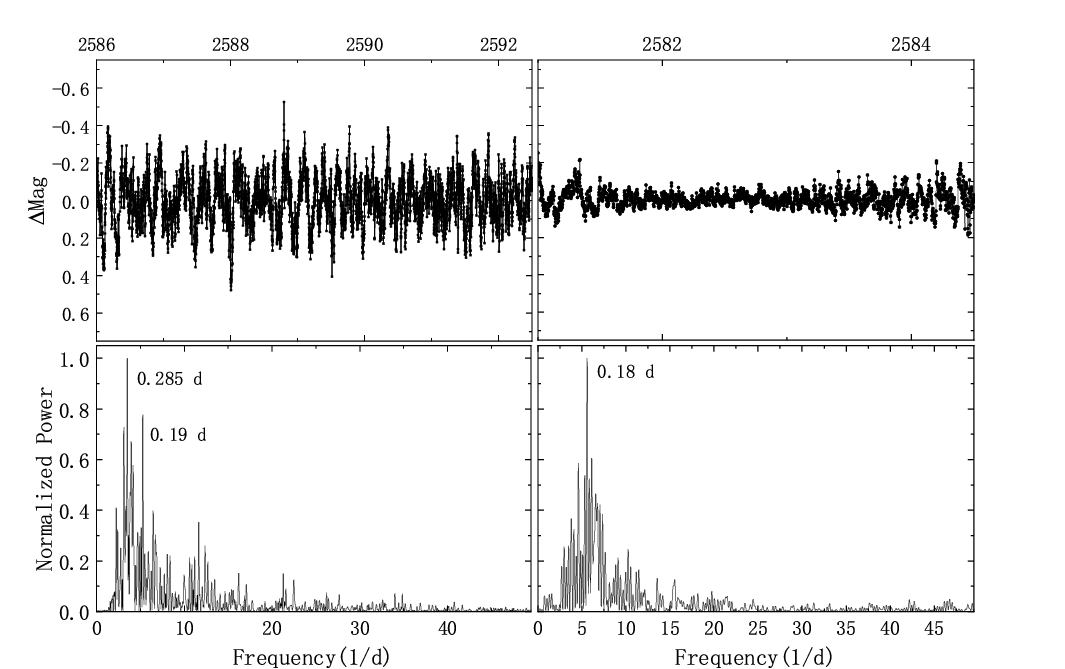}
    \caption{The left and right top panels show the detrended light curve in quiescence and a normal outburst, respectively. The bottom panels show the Lomb-Scargle periodogram derived by using the Corresponding light curve in the top panel.}
    \label{fig:flybh}
\end{figure*}

To investigate QPOs in the light curves, the generalized Lomb-Scargle method \citep{1976Ap&SS..39..447L,1982ApJ...263..835S,1989ApJ...338..277P}
was adopted. The light curves in quiescence from BJD 2459585 to 2459593 and in normal outburst from BJD 2459580 to 2459585 were used to detect the QPOs. Before producing the Lomb-Scargle periodogram, the light curves were detrended by the LOWESS method. In normal outbursts, the trend of the outbursts in the light curve should be removed. In the quiescence, there is still the overall trend in the light curve that needs to be removed, as seen in the top panel of Fig. \ref{fig:overall}. The periodograms and detrended light curves are shown in Fig. \ref{fig:flybh}. The periodgrams show two peaks at 0.285 and 0.19 days in quiescence and the strongest peak at 0.18 days in a normal outburst. We also conduct QPO detection on other quiescence and outbursts of these light curves, and the results are basically the same. At the time of the superoutburst, the QPO was not detected. The amplitude of the QPO in quiescence is significantly higher than that in normal outbursts. The QPO signal of 0.18 days in the normal outbursts is close to the 0.19-day signal in quiescence. Their mechanisms of generation may be the same, and both quiescence and normal outbursts are present in this system. 
The QPO of 0.285 days exists only at the time of quiescence, which is approximately 4 times the orbital period. The mechanism by which they arise is currently unclear. These QPOs mainly appear in quiescence and disappear in outbursts, their timescales are much longer than those of previous QPOs, and their amplitudes are larger.

\section{Conclusion}
Superoutbursts of SU UMa have been observed many times (e.g., \citet{1984PASP...96..988S,1990AJ....100..226U,2010PASJ...62.1525K}), but only a partial light curve of a superoutburst has been obtained, and the evolution of the superhump during the superoutburst cannot be well studied.
It is difficult to observe a complete light curve of a superoutburst, and due to the TESS mission, we were able to obtain such high-precision and continuous light curves. 
Based on the light curves of SU UMa observed by TESS, we analyzed the evolution of superhumps during the 2019 December superoutburst, which shows typical stages A, B, and C. We derived a mean superhump period of 0.079164(9) days and a stage A superhump period of 0.08013(4) days. The mass ratio $q = 0.137 (1)$ is estimated by the stage A superhump method. The donor mass and white dwarf mass are estimated as $M_2 = 0.138(1) M_\odot $ and $M_1 = 1 M_\odot$, respectively.  
 
We also investigated the period evolution of the superhump minimum by the O - C method, and the distribution of O - C shows a parabolic shape with an opening downwards, which is slightly different from that of the superhump maxima O - C diagram. A periodic oscillation with a period of 2.17(9) days exists at the top of the parabola. Using the relationship among $P_{prec}$, $P_{sh}$ and $P_{orb}$, we verified that 2.17(9) days coincides with the precession cycle and that the 2.17(9)-day oscillation is related to the precession. This shows that the O - C method may be a new tool to obtain the period of precession.
 
QPOs were found in quiescence and normal outbursts, but not in the superoutburst. An ultralong period of 0.18 days of QPO was found at the time of a normal outburst, and two ultralong periods of QPOs of 0.19 days and 0.285 days were found in quiescence. These QPOs are unexpected because they have longer periods and higher amplitudes and are more pronounced in quiescence.

\section*{Acknowledgements}

This work was supported by the National Natural Science Foundation of China  (Nos. 11933008, 12103084), and the basic research project of Yunnan Province (Grant No. 202301AT070352), and the Natural Science Foundation of Anhui Province (2208085QA23). This paper includes data collected by the TESS mission, which are publicly available from the Mikulski Archive for Space Telescopes  (MAST). Funding for the TESS mission is provided by the NASA Science Mission Directorate.

%% The Appendices part is started with the command \appendix;
%% appendix sections are then done as normal sections
%\appendix

%\section{Appendix title 1}
%% \label{}

%\section{Appendix title 2}
%% \label{}

%% If you have bibdatabase file and want bibtex to generate the
%% bibitems, please use
%%
\bibliographystyle{elsarticle-harv} 
\bibliography{example}

%% else use the following coding to input the bibitems directly in the
%% TeX file.

%%\begin{thebibliography}{00}

%% \bibitem[Author(year)]{label}
%% For example:

%% \bibitem[Aladro et al.(2015)]{Aladro15} Aladro, R., Martín, S., Riquelme, D., et al. 2015, \aas, 579, A101

%%\end{thebibliography}

\end{document}